\documentclass[11pt,letterpaper]{article}

\usepackage[margin=1in]{geometry}

\usepackage{newtxtext,newtxmath}

\usepackage{amsmath}
\usepackage{graphicx}
\usepackage{booktabs}
\usepackage{tabularx}

\usepackage{caption}
\usepackage{titlesec}
\usepackage{setspace}

\usepackage{nomencl}
\usepackage{array}
\usepackage{url}
\usepackage{subcaption}

\usepackage[hang,flushmargin]{footmisc}
\titleformat{\section}
  {\normalfont\normalsize\bfseries\MakeUppercase}
  {\thesection.}{1em}{}

\titleformat{\subsection}
  {\normalfont\normalsize\bfseries}
  {\thesubsection.}{1em}{}

\titleformat{\subsubsection}
  {\normalfont\normalsize\bfseries}
  {\thesubsubsection.}{1em}{}

\titlespacing*{\section}{0pt}{\baselineskip}{\baselineskip}
\titlespacing*{\subsection}{0pt}{\baselineskip}{\baselineskip}
\titlespacing*{\subsubsection}{0pt}{\baselineskip}{\baselineskip}

\begin{document}

{\centering 
{\fontsize{14pt}{14pt}\selectfont \bfseries Super-Resolution Initialization of High-Fidelity CFD Simulations for Pebble-Bed Reactors}

\vspace{\baselineskip}

\textbf{Guilherme Gottems\footnote{\noindent gkg5288@psu.edu}, Vasil Ivakimov, Luiz Aldeia Machado, Mahmoud Yaseen, Tri Nguyen, and Elia Merzari} \\
Department of Nuclear Engineering \\
The Pennsylvania State University \\
University Park, PA, USA \\
gkg5288@psu.edu; vai5027@psu.edu; lca5209@psu.edu; mqy5284@psu.edu; nguyen.tri@psu.edu; ebm5351@psu.edu

\vspace{\baselineskip}

\textbf{Riccardo Balin, Dillon Shaver, Filippo Simini, Bethany Lusch, Venkatram Vishwanath, Haomin Yuan, Misun Min, Ramesh Balakrishnan, and Jun Fang} \\
Argonne National Laboratory \\
Lemont, IL, USA \\
rbalin@anl.gov; dshaver@anl.gov; fsimini@anl.gov; blusch@anl.gov; venkat@anl.gov; hyuan@anl.gov; mmin@mcs.anl.gov; bramesh@anl.gov; fangj@anl.gov
\vspace{\baselineskip}

\textbf{Paul Fischer} \\
Department of Computer Science \\
University of Illinois Urbana-Champaign \\
Urbana, IL, USA \\
fischerp@illinois.edu


}


\textbf{ABSTRACT}

High-order CFD simulations provide detailed resolution of the heterogeneous interstitial flow in pebble-bed reactors, but their computational cost is high, especially during the initial flow-development period required to reach statistically stationary conditions. This work investigates the use of a Super-Resolution Graph Neural Network (SR-GNN) to improve the initialization of high-order NekRS simulations. Lower-order $P=2$ velocity fields are used as inputs to reconstruct higher-order representations, which are then used as initial conditions for $P=7$ restart simulations. The approach is evaluated using a 146-pebble bed at $Re=1000$, $Re=2500$, and $Re=5000$, with pressure-drop convergence used as the main figure of merit. The SR-GNN models were trained using paired low- and high-order snapshots and were first evaluated through qualitative inference comparisons. High-order restart simulations showed that, for $Re=1000$ and $Re=2500$, the SR-GNN initialized cases produced pressure-drop histories similar to direct restarts from true $P=2$ fields. For $Re=5000$, however, the super-resolved field restart approached the statistically stationary $P=7$ pressure-drop range faster than both the direct $P=2$ restart and the reference $P=7$ simulation initialized from a uniform velocity field. The trained $Re=5000$ model was also applied to a larger 1568-pebble bed, demonstrating qualitative applicability of the workflow to a significantly larger packed-bed geometry. These results indicate that SR-GNN-based initialization is a promising strategy for reducing high-order flow-development cost, while also motivating further work on broader Reynolds-number and geometry generalization.

\vspace{2\baselineskip}

{\raggedleft \textbf{KEYWORDS} \\ 
CFD, NekRS, Super Resolution, Graph Neural Network, Machine Learning \par}

\vspace{\baselineskip}

\section*{Nomenclature}

\textbf{Symbols}

\begin{tabularx}{\textwidth}{@{}lXl@{}}
\toprule
Symbol & Description & Unit \\
\midrule
$D$              & Cylinder diameter                              & m \\
$d_p$            & Pebble diameter and reference length           & m \\
$H$              & Height of the active packed region             & m \\
$n_e$            & Number of spectral elements                    & -- \\
$P$              & Polynomial order                               & -- \\
$p$              & Dimensional pressure                           & Pa \\
$p^*$            & Nondimensional pressure                        & -- \\
$Re$             & Reynolds number                                & -- \\
$t$              & Dimensional time                               & s \\
$t^*$            & Nondimensional time                            & -- \\
$U_{in}$         & Dimensional reference inlet velocity           & m\,s$^{-1}$ \\
$U_{in}^*$       & Nondimensional inlet velocity                  & -- \\
$\mathbf{u}$     & Dimensional velocity vector                    & m\,s$^{-1}$ \\
$\mathbf{u}^*$   & Nondimensional velocity vector                 & -- \\
$\mathbf{x}$     & Dimensional position vector                    & m \\
$\mathbf{x}^*$   & Nondimensional position vector                 & -- \\
$\Delta p^*$     & Nondimensional pressure drop across the bed    & -- \\
$p_{in}^*$       & Nondimensional inlet pressure                  & -- \\
$p_{out}^*$      & Nondimensional outlet pressure                 & -- \\
$\mu$            & Dynamic viscosity                              & Pa\,s \\
$\rho$           & Fluid density                                  & kg\,m$^{-3}$ \\
\bottomrule
\end{tabularx}

\vspace{\baselineskip}

\textbf{Acronyms}

\begin{tabularx}{\textwidth}{@{}lX@{}}
\toprule
Acronym & Definition \\
\midrule
CFD      & Computational fluid dynamics \\
DNS      & Direct numerical simulation \\
DOE      & U.S. Department of Energy \\
FT       & Flow-through time \\
GLL      & Gauss--Lobatto--Legendre \\
GPU      & Graphics processing unit \\
INCITE   & Innovative and Novel Computational Impact on Theory and Experiment \\
LES      & Large eddy simulation \\
ML       & Machine learning \\
MSE      & Mean squared error \\
PIV      & Particle image velocimetry \\
SR-GNN   & Super-Resolution Graph Neural Network \\
TRISO    & Tri-structural isotropic \\
\bottomrule
\end{tabularx}

\vspace{\baselineskip}

\section{Introduction}

Pebble-bed reactors are an advanced nuclear reactor concept that employs spherical fuel elements known as pebbles. Each pebble contains thousands of tri-structural isotropic (TRISO) fuel particles, typically composed of a uranium-based fuel kernel surrounded by multiple ceramic and carbon coating layers. This fuel design enhances reactor safety by improving fission-product retention, increasing resistance to oxidation and corrosion, and allowing operation at very high temperatures.

Inside the reactor core, the pebbles are randomly packed, producing a complex thermal-hydraulic system in which the interstitial flow is highly heterogeneous. Accurate prediction of the fluid dynamics and heat transfer behavior in this type of reactor remains challenging and is particularly important for safety analysis. High-fidelity simulations have shown preferential flow paths, strong heterogeneity in the interstitial fluid motion, and pronounced near-wall effects caused by the reduced local packing density near confining boundaries~\cite{david_wall_channeling}. These features highlight the importance of resolving the relevant flow scales within the packed bed.

However, the computational cost associated with simulating a full commercial-scale pebble-bed core remains prohibitively high, even on modern leadership-class supercomputers. In addition, turbulent CFD simulations are commonly initialized from idealized flow fields, such as uniform velocity profiles, and must be advanced through an initial transient before a statistically stationary state is reached and meaningful time-averaged quantities can be collected. This flow-development period can represent a significant fraction of the total computational cost, especially for high-order simulations of complex packed-bed geometries.

Reduced order model-assisted approaches and machine learning (ML) methods provide promising pathways to reduce this cost. In particular, ML-based super-resolution methods can be used to bridge lower-resolution simulations and high-fidelity representations by reconstructing missing flow scales from cheaper numerical solutions. In this context, the present work investigates the use of a Super-Resolution Graph Neural Network (SR-GNN) to improve the initialization of high-fidelity CFD simulations of turbulent flow in packed-bed geometries relevant to pebble-bed reactors. The objective is to provide a higher-quality restart field that reduces the time to solution and the respective computational cost of high-order simulations.

\section{Methodology}

\subsection{High-order CFD solver: NekRS}\label{subsec:nekrs}

The CFD calculations in this work are performed with NekRS, a GPU-accelerated high-order spectral element solver for incompressible flow~\cite{nekrs}. The governing equations are solved in nondimensional form. The nondimensional variables are defined as

\begin{equation}
\mathbf{x}^* = \frac{\mathbf{x}}{d_p},
\qquad
\mathbf{u}^* = \frac{\mathbf{u}}{U_{in}},
\qquad
t^* = \frac{t U_{in}}{d_p},
\qquad
p^* = \frac{p}{\rho U_{in}^2},
\end{equation}

where $d_p$ is the pebble diameter and $U_{in}$ is the prescribed
dimensional inlet velocity. Using these definitions, the nondimensional incompressible Navier--Stokes equations are

\begin{align}
\mathbf{\nabla}^* \cdot \mathbf{u^*} &= 0, \\
 \frac{\partial \mathbf{u}^*}{\partial t^*} + \mathbf{u}^* \cdot \mathbf{\nabla}^* \mathbf{u}^* &= -\mathbf{\nabla}^* p^* + \frac{1}{Re} \nabla^{*2} \mathbf{u}^*
\end{align}

where the Reynolds number is defined as

\begin{equation}
Re = \frac{\rho U_{in}d_p}{\mu}.
\end{equation}

With this nondimensionalization, the prescribed inlet velocity is
$U_{in}^*=1$.

The turbulent flow simulations considered in this work are performed using the large eddy simulation (LES) approach.

NekRS uses the spectral element method, in which the computational domain is decomposed into high-order hexahedral elements. Within each element, the solution is represented on Gauss--Lobatto--Legendre (GLL) points using polynomial basis functions. For a polynomial order $P$, each element contains $(P+1)^3$ nodal points in three dimensions, giving approximately $n_e(P+1)^3$ local solution points before accounting for shared element interfaces, where $n_e$ is the number of spectral elements. This formulation enables $p$-refinement, where the polynomial order is increased to improve the resolution of small-scale flow features without changing the underlying mesh topology.

This property is central to the present work. Lower-order simulations, such as $P=2$, are significantly cheaper than higher-order calculations but under-resolve part of the turbulent flow structure. Higher-order simulations, such as $P=7$, provide improved spatial resolution but require greater computational cost, particularly during the initial flow-development period before statistically stationary behavior is reached. The super-resolution approach investigated here uses lower-order NekRS fields as inputs to construct improved initial conditions for subsequent high-order restart simulations.

In this study, NekRS is used both to generate high-order reference solutions and to produce lower-order fields used for SR-GNN training, inference, and restart assessment. The simulations are integrated with the NekRS-ML workflow, which provides access to the spectral element mesh, solution fields, and partitioning information required to construct graph-based machine-learning inputs and to transfer the reconstructed fields back into NekRS for high-order restart calculations.

\subsubsection{NekRS V\&V}

The high-order spectral element methodology used in NekRS and its predecessor Nek5000 has been extensively assessed for pebble-bed flow applications. Fick et al.~\cite{fick_verification} performed direct numerical simulations (DNS) of flow through an idealized extended face-centered-cubic pebble-bed configuration and evaluated the numerical resolution using multiple polynomial orders. The grid spacing and time step size were compared with the Kolmogorov length and time scales, respectively, and good agreement was observed among the sufficiently resolved discretizations for the velocity statistics and turbulent spectra. The results were also cross-verified against available quasi-DNS data~\cite{shams_qdns}, demonstrating that the spectral element formulation can adequately resolve the relevant turbulent scales in pebble-bed flows.

Validation was subsequently performed for a randomly packed pebble-bed geometry using experimental data from Nguyen et al.~\cite{nguyen_validation_exp}. Their matched index of refraction facility and time-resolved particle image velocimetry (PIV) measurements provided detailed first- and second-order velocity statistics within the interstitial regions of the packed bed. Yildiz et al.~\cite{yildiz_pb146} reconstructed the experimental packing from measured sphere locations and performed DNS using Nek5000. The numerical resolution was assessed against the Kolmogorov length scale, while the resulting bed porosity and pressure drop were compared with established packed-bed correlations. The computed velocity profiles and Reynolds stresses were compared directly with the experimental measurements. Overall, good agreement was obtained for both the first- and second-order flow statistics, providing validation of the high-order spectral element methodology for randomly packed pebble-bed flows.

\subsection{Pebble-bed geometries and simulation setup}\label{subsec:simulation_setup}

Two packed-bed geometries are considered in this work: a 146-pebble bed used as the primary training and restart-assessment case, and a larger 1568-pebble bed used to evaluate the applicability of the SR-GNN workflow to a more reactor-relevant configuration. The two cases differ in the number of pebbles, packing characteristics, mesh size, and total number of solution points.

\begin{figure}[h!]
  \centering
  \begin{minipage}[t]{0.5\linewidth}
    \centering
    \includegraphics[width=\linewidth]{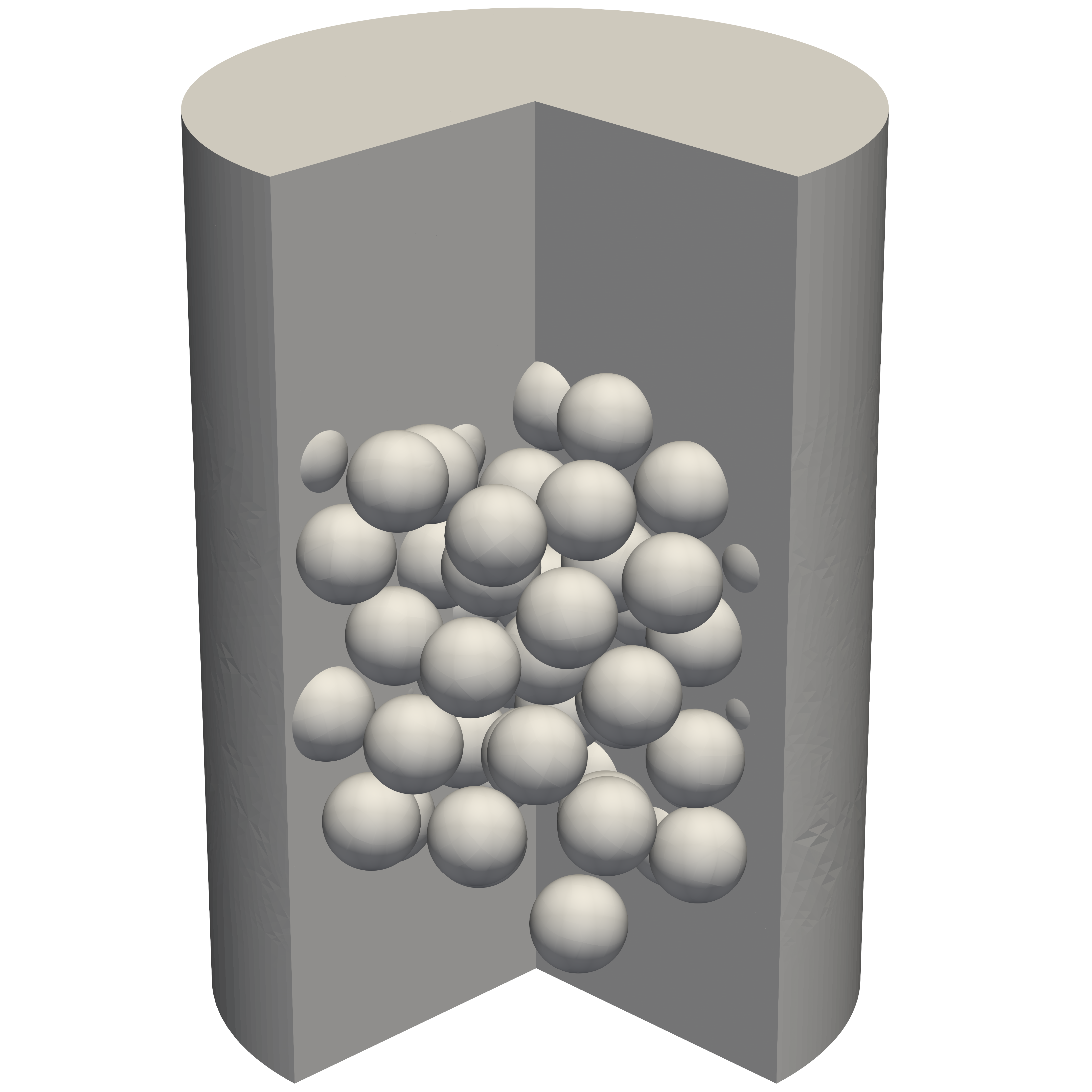}
    \caption{146-pebble case geometry.}
    \label{fig:pb146_geometry}
  \end{minipage}\hfill
  \begin{minipage}[t]{0.5\linewidth}
    \centering
    \includegraphics[width=\linewidth]{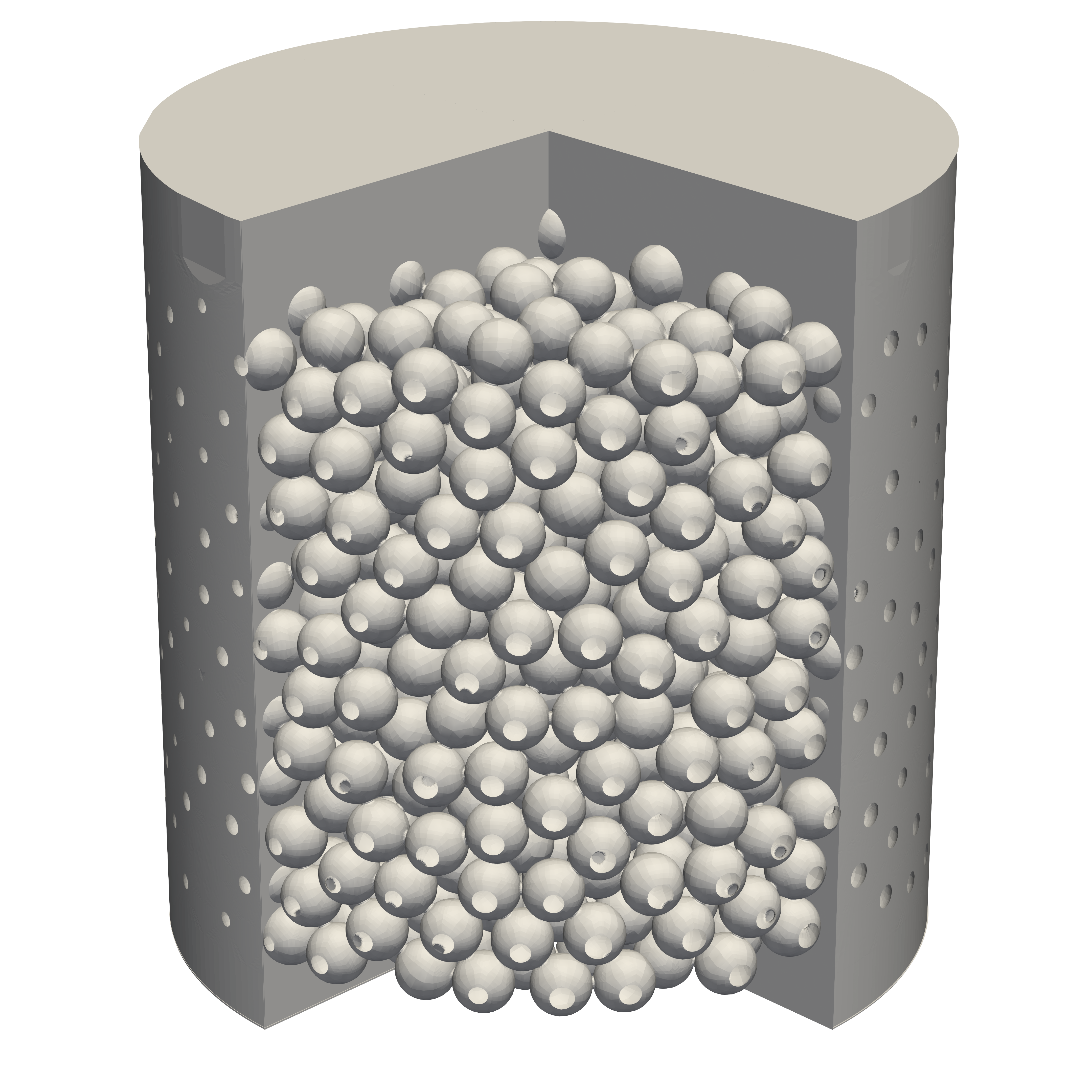}
    \caption{1568-pebble case geometry.}
    \label{fig:pb1568_geometry}
  \end{minipage}
\end{figure}

The smaller geometry contains 146 pebbles, and the computational mesh was generated from the experimental configuration used in the validation study of Yildiz et al.~\cite{yildiz_pb146}. The mesh contains 62,132 spectral elements. Simulations with polynomial orders $P=2$ and $P=7$ correspond to approximately 1.7 million and 32 million grid points, respectively. This geometry is used to train the SR-GNN model and to evaluate the effectiveness of the super-resolved field as an initial condition for high-order restart simulations. The computational domain is shown in Fig.~\ref{fig:pb146_geometry}. The cylinder-to-pebble diameter ratio is $D/d_p=14$, the active-core height is approximately $H/d_p=11$, and the total domain height is approximately $20d_p$, where $D$ is the cylinder diameter, $d_p$ is the pebble diameter, and $H$ is the height of the active packed region.

The larger geometry contains 1568 pebbles randomly packed inside a cylindrical container. In this mesh, contacts between neighboring pebbles and between pebbles and the cylinder wall are represented using small chamfers. This treatment avoids extremely narrow gaps near contact points, reducing the computational cost while preserving the main packing structure and porosity characteristics of the bed. The mesh contains approximately 2 million spectral elements. A $P=2$ simulation contains approximately 54 million GLL solution points, while the corresponding $P=7$ discretization would contain approximately 1 billion grid points. This case is therefore used as a preliminary large-geometry demonstration of the SR-GNN inference workflow. The cylinder-to-pebble diameter ratio is $D/d_p=6.5$, the active-core height is approximately $H/d_p=12$, and the total domain height is approximately $14d_p$. The computational domain is shown in Fig.~\ref{fig:pb1568_geometry}.

The configurations considered in this work are summarized in Table~\ref{tab:case_summary}. The 146-pebble geometry is used for SR-GNN training, qualitative inference, and high-order restart assessment at three Reynolds numbers. The 1568-pebble geometry is used only for cross-geometry inference with the model trained on the 146-pebble case at $Re=5000$; no corresponding $P=7$ reference solution or high-order restart calculation is performed for this
larger geometry.

\begin{table}[h!]
\centering
\caption{Summary of the configurations considered and their roles
in the SR-GNN workflow.}
\label{tab:case_summary}
\small
\begin{tabularx}{\textwidth}{@{}ccXX@{}}
\toprule
Pebbles & $Re$ & Velocity fields used or generated
& Role in this work \\
\midrule
146
& 1000
& $P=2$ input, SR-GNN reconstructed $P=7$, and reference $P=7$ from NekRS
& SR-GNN training, qualitative inference, and $P=7$ restart assessment. \\

146
& 2500
& $P=2$ input, SR-GNN reconstructed $P=7$, and reference $P=7$ from NekRS
& SR-GNN training, qualitative inference, and $P=7$ restart assessment. \\

146
& 5000
& $P=2$ input, SR-GNN reconstructed $P=7$, and reference $P=7$ from NekRS
& SR-GNN training, qualitative inference, and $P=7$ restart assessment. \\

1568
& 5000
& $P=2$ from NekRS input and SR-GNN reconstructed $P=7$
& Cross-geometry inference using the model trained on the 146-pebble
case with $Re=5000$. \\
\bottomrule
\end{tabularx}
\end{table}

For both geometries, a uniform upward velocity profile is prescribed at the inlet. No-slip boundary conditions are imposed on the pebble surfaces, chamfer surfaces, and cylinder walls, and a stabilized zero-gradient outflow condition is applied at the outlet.

\subsection{SR-GNN model}\label{subsec:srgnn}

The machine-learning component of this work is based on the Super-Resolution Graph Neural Network (SR-GNN) architecture developed for high-order spectral element flow simulations~\cite{srgnn}. The objective of the model is to reconstruct a higher-order representation of the velocity field from a lower-order input defined on the same spectral element mesh. In the present study, the SR-GNN prediction is not used as a replacement for the CFD solution. Instead, it is used to generate an improved initial condition for subsequent high-order NekRS restart simulations.

The spectral element mesh is represented as a graph in which the nodes correspond to elements and the edges describe local connectivity between neighboring elements. This representation is natural for NekRS simulations because the element topology is preserved when changing the polynomial order. Therefore, a lower-order and a higher-order solution can be associated with the same element-level graph, while the number of solution points inside each element changes according to the polynomial order. The graph structure also allows the model to account for local information exchange between adjacent elements, which is important in packed-bed flows where the interstitial velocity field varies strongly between packed layers.

The SR-GNN operates locally on each spectral element and, depending on the model configuration, on its immediate element neighborhood. Within this local stencil, the model combines coarse-scale information from neighboring elements with fine-scale reconstruction inside each element. This structure is well suited for spectral element simulations because the solution inside each element follows a tensor-product arrangement of GLL points, while the global packed-bed geometry remains unstructured at the element-connectivity level. As a result, the model can learn local mappings between under-resolved and high-order flow structures without requiring a globally structured mesh.

The supervised training datasets are constructed from paired low- and high-order velocity fields. High-order NekRS simulations provide the target fields, while the corresponding lower-order inputs are obtained by interpolating the high-order solution to a lower polynomial representation on the same element topology. This procedure produces consistent pairs of input and target fields for each element, allowing the model to learn the mapping from lower-order to higher-order solution content. Although this approach requires high-order data during training, the trained model can then be applied to lower-order simulations to generate super-resolved fields at a much lower cost than advancing the full high-order simulation through the same flow-development period.

This ability to apply the trained model to fields that were not included in the training set is an important aspect of the SR-GNN approach. Previous demonstrations of the SR-GNN showed that graph-based super-resolution models can recover fine-scale flow structures not only for unseen snapshots, but also for configurations involving changes in geometry and Reynolds number~\cite{srgnn}. These results motivate the present study, where SR-GNN models trained on the 146-pebble geometry are evaluated both through high-order restart calculations in the same geometry and through qualitative inference on the larger 1568-pebble bed. The latter case is used to assess whether the learned local reconstruction can be applied to a larger packed-bed configuration, while recognizing that full validation of acceleration requires comparison against high-order restart simulations.

\subsection{High-order restart and acceleration assessment}\label{subsec:restart}

The main objective of this work is to generate lower-cost, higher-quality initial conditions for high-order CFD simulations using lower-order flow fields. This is achieved by applying the trained SR-GNN model to a lower-order NekRS snapshot that was not used during training. The resulting super-resolved velocity field is then used as an initial condition for a high-order restart simulation.

The quality of the SR-GNN-generated initial condition is assessed using the evolution of the total pressure drop across the packed bed. The pressure drop is calculated as

\begin{equation}
    \Delta p^* = p_{in}^* - p_{out}^*
\end{equation}

where $\Delta p^*$ is the total pressure drop, and $p_{in}^*$ and $p_{out}^*$ are the area-averaged pressures at the inlet and outlet, respectively. Pressure drop is used as the main figure of merit because it represents the global hydraulic response of the packed bed and is directly related to the development of the interstitial velocity field. Therefore, convergence of $\Delta p^*$ toward the reference high-order value provides a practical measure of whether the restarted simulation has approached the target high-order flow behavior.

For this comparison, the flow is first advanced using a lower-order simulation for one flow-through (FT). The corresponding lower-order velocity field is then used in two ways. First, it is directly used as the initial condition for a high-order restart simulation. Second, it is passed through the SR-GNN model to generate a super-resolved velocity field, which is then used to initialize a separate high-order restart simulation. The two high-order simulations are advanced in parallel and compared through their pressure-drop histories.

The desired outcome is that the combined cost of the lower-order flow-development simulation, SR-GNN inference, and high-order restart is lower than the cost of reaching the same pressure-drop behavior using a high-order simulation initialized directly from the lower-order field. This approach is also intended to reduce the high-order flow-development time compared with starting the simulation from an artificial initial condition, such as a uniform velocity field throughout the domain.

\section{Results and Discussion}\label{sec:results}
\subsection{146-pebble bed}

The 146-pebble case was used to train separate SR-GNN models for three flow conditions: $Re=1000$, $Re=2500$, and $Re=5000$. For each training case, five paired low- and high-order snapshots were used to represent decorrelated turbulent flow states. The snapshots were extracted after two FTs from the beginning of the simulation and were separated by two convective time units.

\begin{figure}[h!]
    \centering

    \begin{subfigure}[t]{0.33\linewidth}
        \centering
        \includegraphics[width=\linewidth]{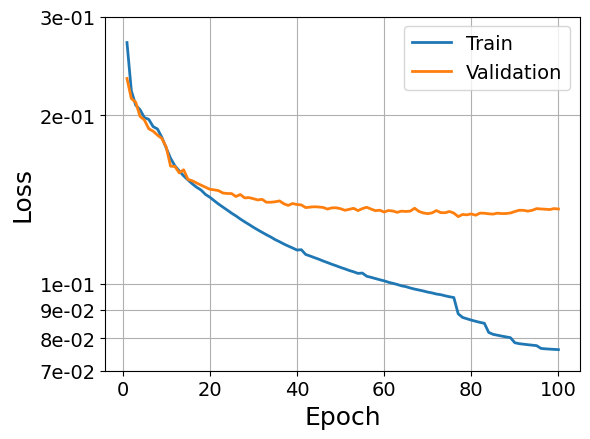}
        \caption{$Re=1000$}
        \label{fig:loss_re1k}
    \end{subfigure}
    \hfill
    \begin{subfigure}[t]{0.33\linewidth}
        \centering
        \includegraphics[width=\linewidth]{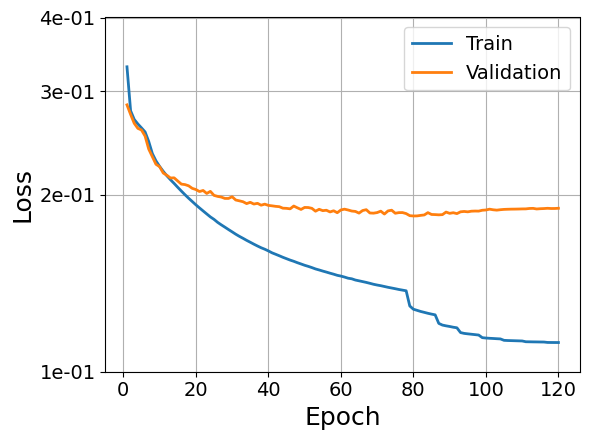}
        \caption{$Re=2500$}
        \label{fig:loss_re2k5}
    \end{subfigure}
    \hfill
    \begin{subfigure}[t]{0.33\linewidth}
        \centering
        \includegraphics[width=\linewidth]{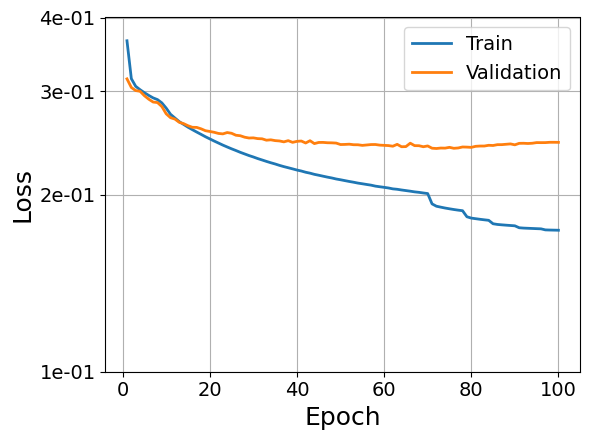}
        \caption{$Re=5000$}
        \label{fig:loss_re5k}
    \end{subfigure}
    
    \caption{Training loss for the SR-GNN models trained on the 146-pebble geometry at different Reynolds numbers.}
    \label{fig:training_losses}
\end{figure}

Figure~\ref{fig:training_losses} shows the mean squared error (MSE) training loss for the three Reynolds-specific SR-GNN models. In all cases, the loss decreases during training, indicating that the model is able to learn the correlation between the lower-order input fields and the corresponding high-order targets for the 146-pebble geometry. 

The models for $Re=1000$ and $Re=5000$ were trained for 100 epochs, while the model for $Re=2500$ was trained for 120 epochs due to slower convergence. For all three Reynolds numbers, the checkpoint used for inference was selected based on the minimum validation MSE to reduce the risk of overfitting. The cases with lower Reynolds number reached lower MSE values than the $Re=5000$ case, which is consistent with the reduced level of small-scale turbulent structures at lower Reynolds numbers.

After training, inference was first evaluated qualitatively using paired snapshots from the 146-pebble simulations. For these visual comparisons, the lower-order input fields were obtained by interpolating the corresponding high-order snapshots to $P=2$. This procedure provides paired low- and high-order fields representing the same instantaneous turbulent state, allowing direct comparison between the lower-order input, the SR-GNN reconstruction, and the high-order target. These interpolated $P=2$ fields are used only for qualitative inference assessment. The restart simulations discussed later use independently advanced, true $P=2$ simulations extracted after one FT of flow development.

\begin{figure}[h!]
    \centering
    \includegraphics[width=1.0\linewidth]{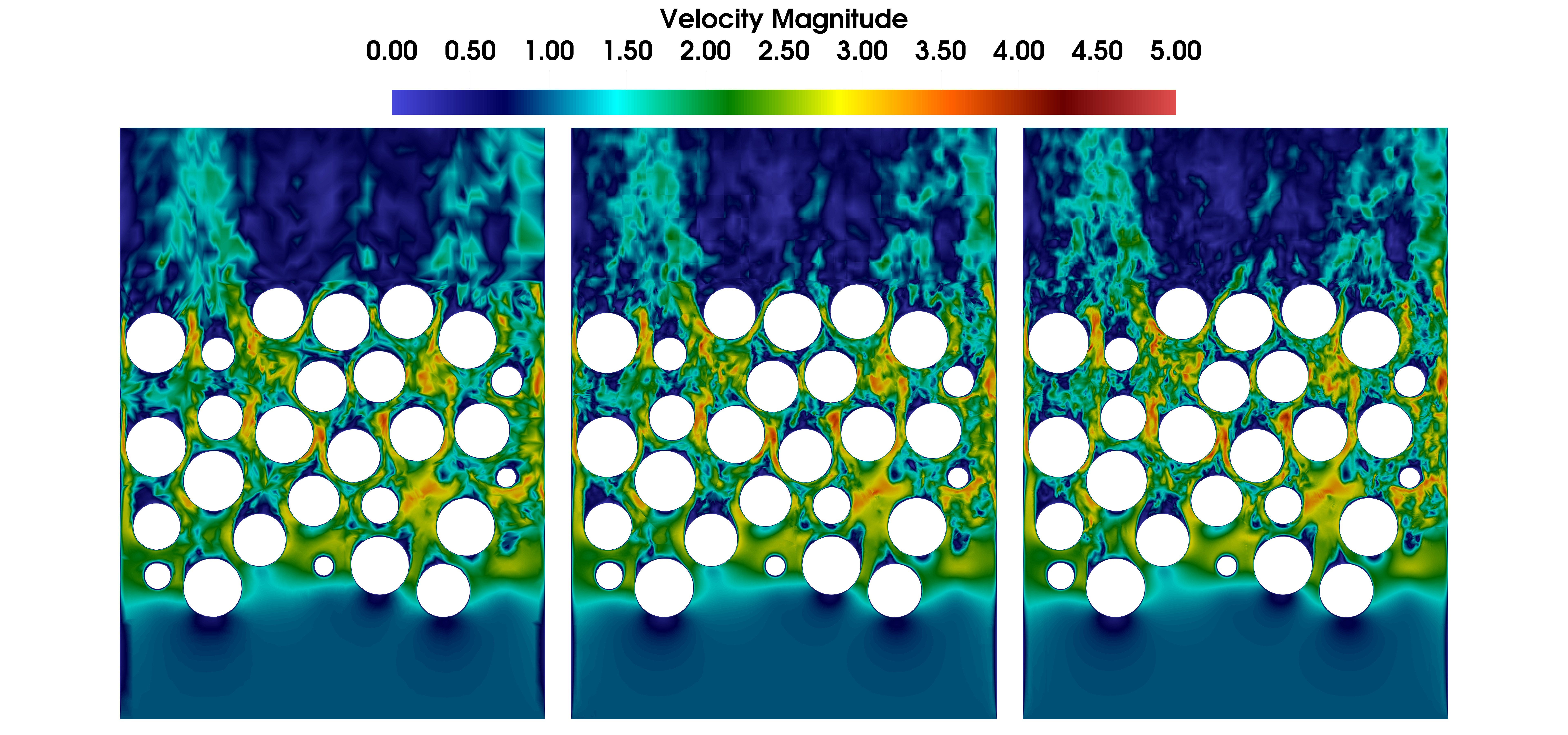}
    \caption{Qualitative SR-GNN inference comparison of the nondimensional velocity magnitude for the 146-pebble bed at $Re=1000$, showing the lower-order input field ($P=2$), the reconstructed field ($P=2 \rightarrow P=7$), and the reference high-order field ($P=7$).}
    \label{fig:inference_re1k}
\end{figure}

\begin{figure}[h!]
    \centering
    \includegraphics[width=1.0\linewidth]{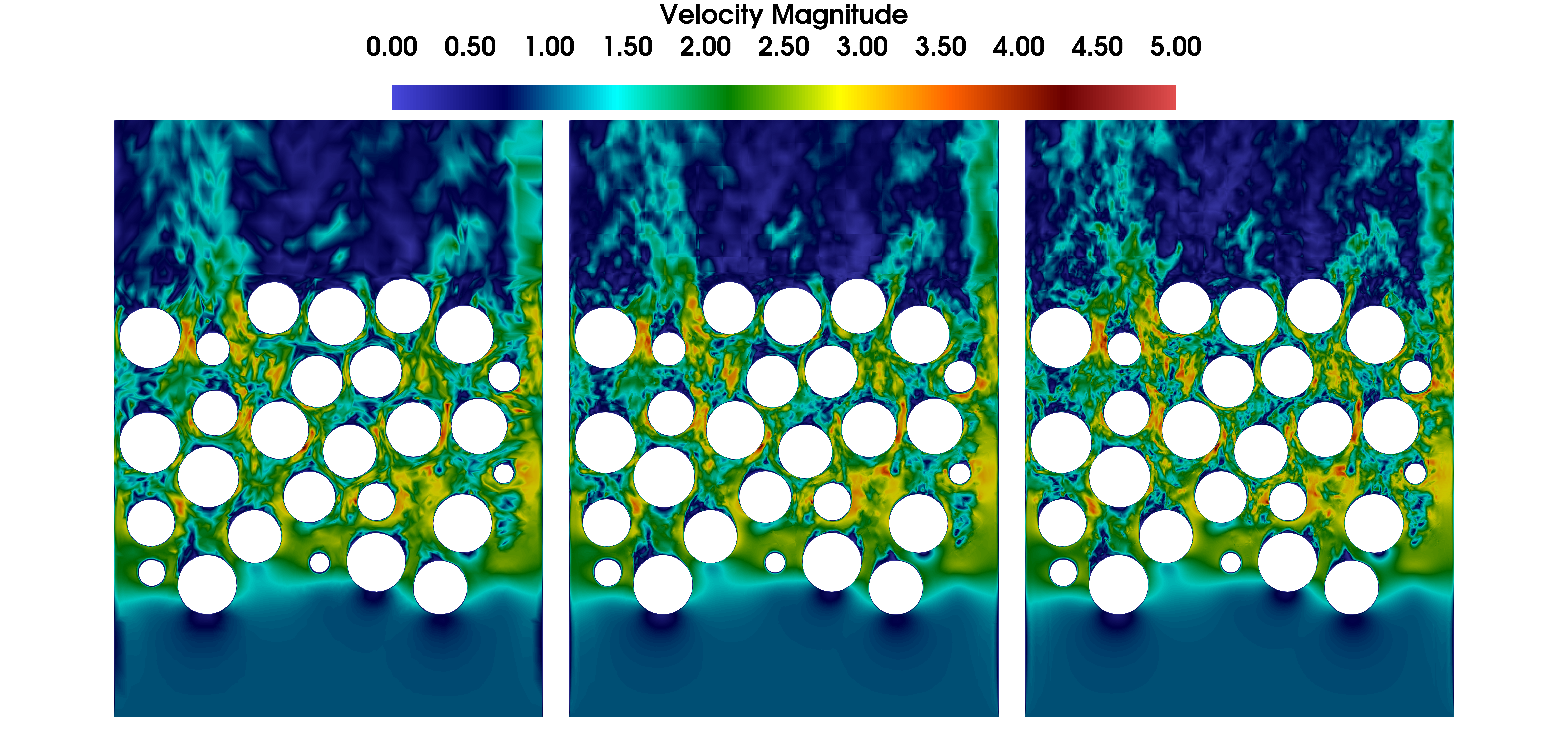}
    \caption{Qualitative SR-GNN inference comparison of the nondimensional velocity magnitude for the 146-pebble bed at $Re=2500$, showing the lower-order input field ($P=2$), the reconstructed field ($P=2 \rightarrow P=7$), and the reference high-order field ($P=7$).}
    \label{fig:inference_re2k5}
\end{figure}

\begin{figure}[h!]
    \centering
    \includegraphics[width=1.0\linewidth]{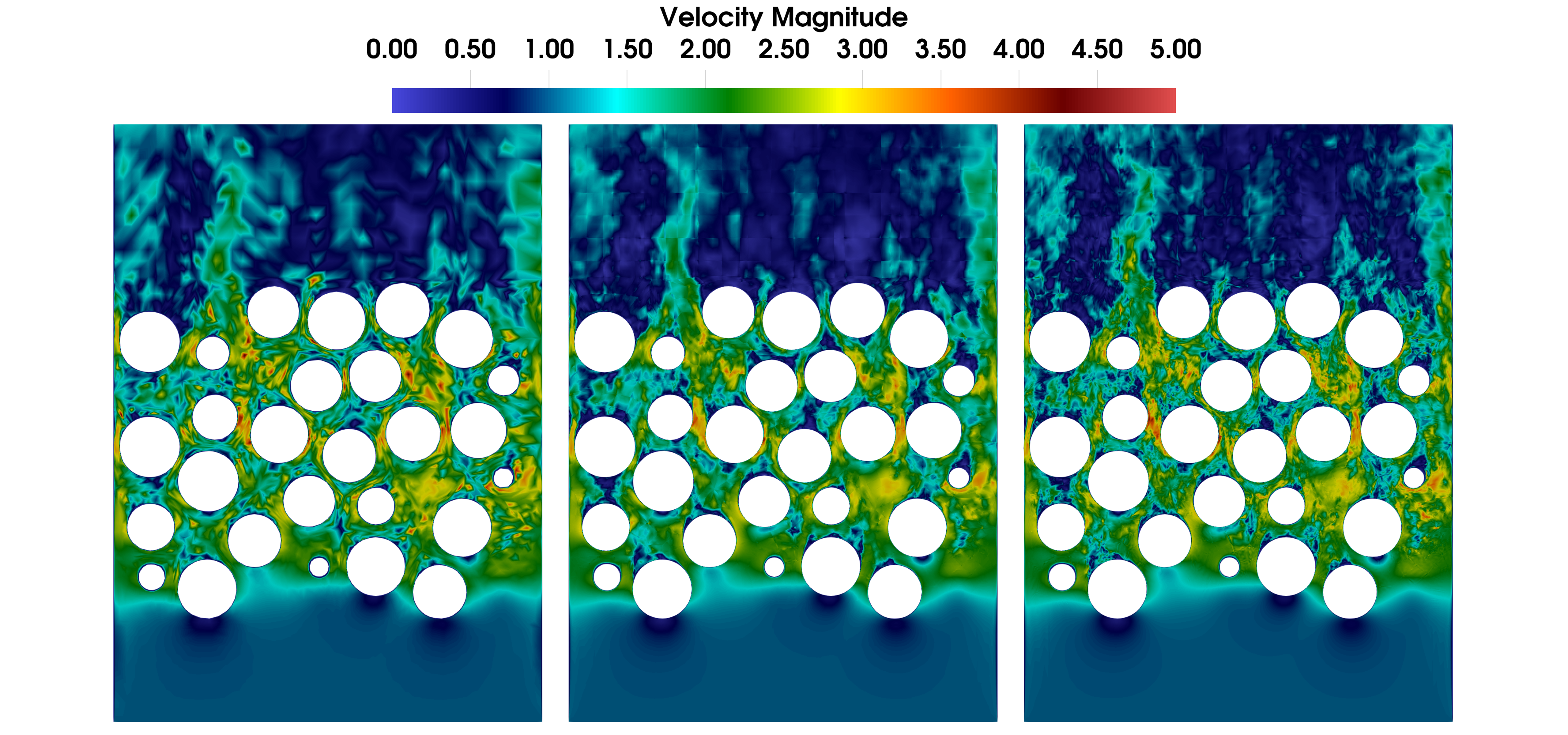}
    \caption{Qualitative SR-GNN inference comparison of the nondimensional velocity magnitude for the 146-pebble bed at $Re=5000$, showing the lower-order input field ($P=2$), the reconstructed field ($P=2 \rightarrow P=7$), and the reference high-order field ($P=7$).}
    \label{fig:inference_re5k}
\end{figure}

Figures~\ref{fig:inference_re1k},~\ref{fig:inference_re2k5}, and~\ref{fig:inference_re5k} compare the nondimensional velocity magnitude, obtained from the interpolated lower-order input field ($P=2$), the SR-GNN reconstructed field ($P=2 \rightarrow P=7$), and the reference high-order solution ($P=7$) for the three Reynolds numbers.

The qualitative comparisons indicate whether the SR-GNN model reconstructs high-order flow structures from the lower-order input fields. However, visual agreement alone is not sufficient to determine whether the reconstructed fields provide better initial conditions for high-order simulations. Therefore, the following section evaluates the SR-GNN predictions through high-order restart simulations and compares the resulting pressure-drop histories.

\subsubsection{High-Order Restart Assessment}

The qualitative inference comparisons indicate whether the SR-GNN model can reconstruct high-order flow structures from lower-order inputs. However, the primary objective of this work is to determine whether the reconstructed fields provide improved initial conditions for high-order NekRS simulations. To evaluate this, high-order restart simulations were performed for each Reynolds number using two different initial conditions extracted after one FT of lower-order flow development.

For each case, a true $P=2$ simulation was first advanced for one FT. The resulting velocity field was then used in two ways. First, it was directly interpolated to initialize a $P=7$ restart simulation. Second, it was passed through the SR-GNN model trained at the corresponding Reynolds number, and the reconstructed field was used to initialize a separate $P=7$ restart simulation. These two restart calculations were compared against a reference $P=7$ simulation initialized from a uniform velocity field and advanced from the beginning of the flow-development process.

\begin{figure}[h!]
    \centering
    \includegraphics[width=0.8\linewidth]{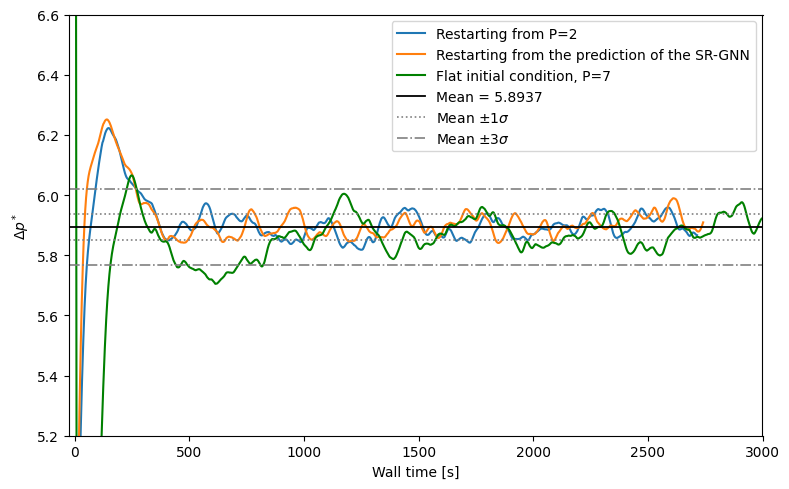}
    \caption{Pressure-drop evolution for the 146-pebble bed at $Re=1000$. The curves compare the reference $P=7$ simulation initialized from a uniform velocity field, the $P=7$ restart initialized from a true $P=2$ field after one FT, and the $P=7$ restart initialized from the SR-GNN reconstructed field.}
    \label{fig:restart_re1k}
\end{figure}

\begin{figure}[h!]
    \centering
    \includegraphics[width=0.8\linewidth]{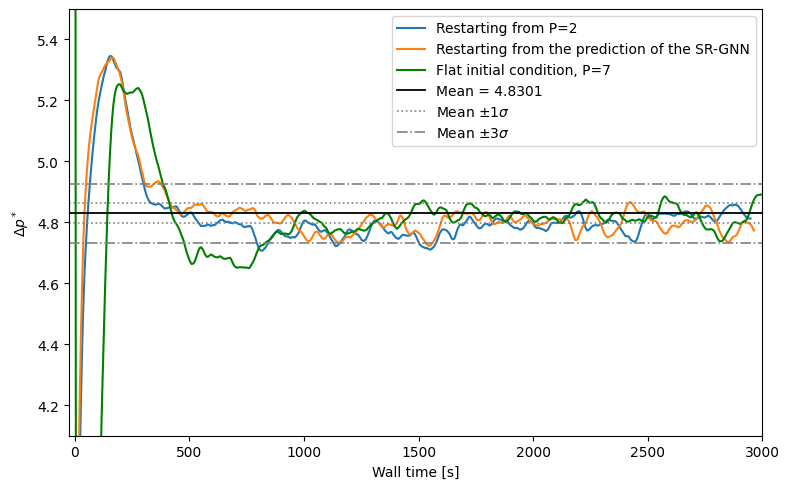}
    \caption{Pressure-drop evolution for the 146-pebble bed at $Re=2500$. The curves compare the reference $P=7$ simulation initialized from a uniform velocity field, the $P=7$ restart initialized from a true $P=2$ field after one FT, and the $P=7$ restart initialized from the SR-GNN reconstructed field.}
    \label{fig:restart_re2k5}
\end{figure}

\begin{figure}[h!]
    \centering
    \includegraphics[width=0.8\linewidth]{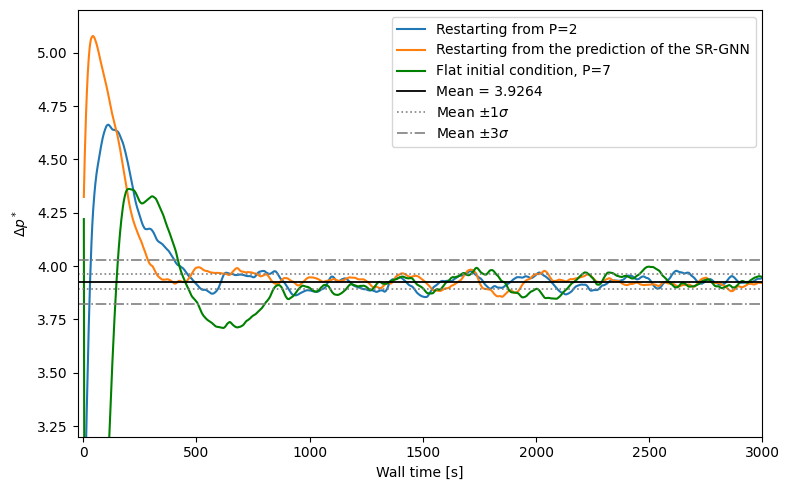}
    \caption{Pressure-drop evolution for the 146-pebble bed at $Re=5000$. The curves compare the reference $P=7$ simulation initialized from a uniform velocity field, the $P=7$ restart initialized from a true $P=2$ field after one FT, and the $P=7$ restart initialized from the SR-GNN reconstructed field.}
    \label{fig:restart_re5k}
\end{figure}

The total pressure drop across the bed was used as the main figure of merit. Figures~\ref{fig:restart_re1k},~\ref{fig:restart_re2k5}, and~\ref{fig:restart_re5k} show the pressure-drop evolution for the reference $P=7$ calculation, the $P=7$ restart from the true $P=2$ field, and the $P=7$ restart from the SR-GNN reconstructed field for $Re=1000$, $Re=2500$, and $Re=5000$, respectively. The reference mean pressure drop and standard-deviation bands were computed from statistically developed $P=7$ simulations over a sampling window of 3 FTs. These quantities provide a reference range for evaluating how quickly each restart approaches the statistically stationary high-order pressure-drop behavior.

In these plots, the wall time is measured from the beginning of each high-order restart and therefore compares the quality of the restart initial condition rather than the total end-to-end computational cost.

For $Re=1000$ and $Re=2500$, Figs.~\ref{fig:restart_re1k} and~\ref{fig:restart_re2k5} show that the SR-GNN reconstructed field does not significantly change the pressure-drop evolution relative to the direct $P=2$ restart. In both cases, the pressure-drop histories from the two restart strategies are nearly identical, indicating that the super-resolved field does not provide a measurable improvement in the global hydraulic convergence for these Reynolds numbers.

A different behavior is observed for $Re=5000$, shown in Fig.~\ref{fig:restart_re5k}. In this case, the SR-GNN-initialized restart exhibits a larger initial pressure-drop excursion, but subsequently approaches the statistically stationary $P=7$ reference range faster than both the reference calculation initialized from a uniform velocity field and the direct restart from the true $P=2$ field. This result indicates that, for the highest Reynolds number considered, the super-resolved velocity field provides a more effective high-order initial condition in terms of pressure-drop convergence.

The Reynolds-number dependence of the restart performance may be associated with differences in the dominant flow structures and with the ability of an element-localized model to reconstruct the missing high-order content from the lower-order input. At lower Reynolds numbers, the direct $P=2$ field may already provide a sufficiently accurate representation of the global pressure-drop behavior, making the effect of super-resolution difficult to observe using this integral metric. At $Re=5000$, where smaller-scale turbulent structures are more relevant, the additional high-order content introduced by the SR-GNN appears to improve the initial condition for the subsequent $P=7$ simulation.

\subsection{Extrapolation to a larger geometry: 1,568-pebble bed}

\begin{figure}[h!]
    \centering

    \begin{subfigure}[t]{0.46\linewidth}
        \centering
        \includegraphics[width=\linewidth]{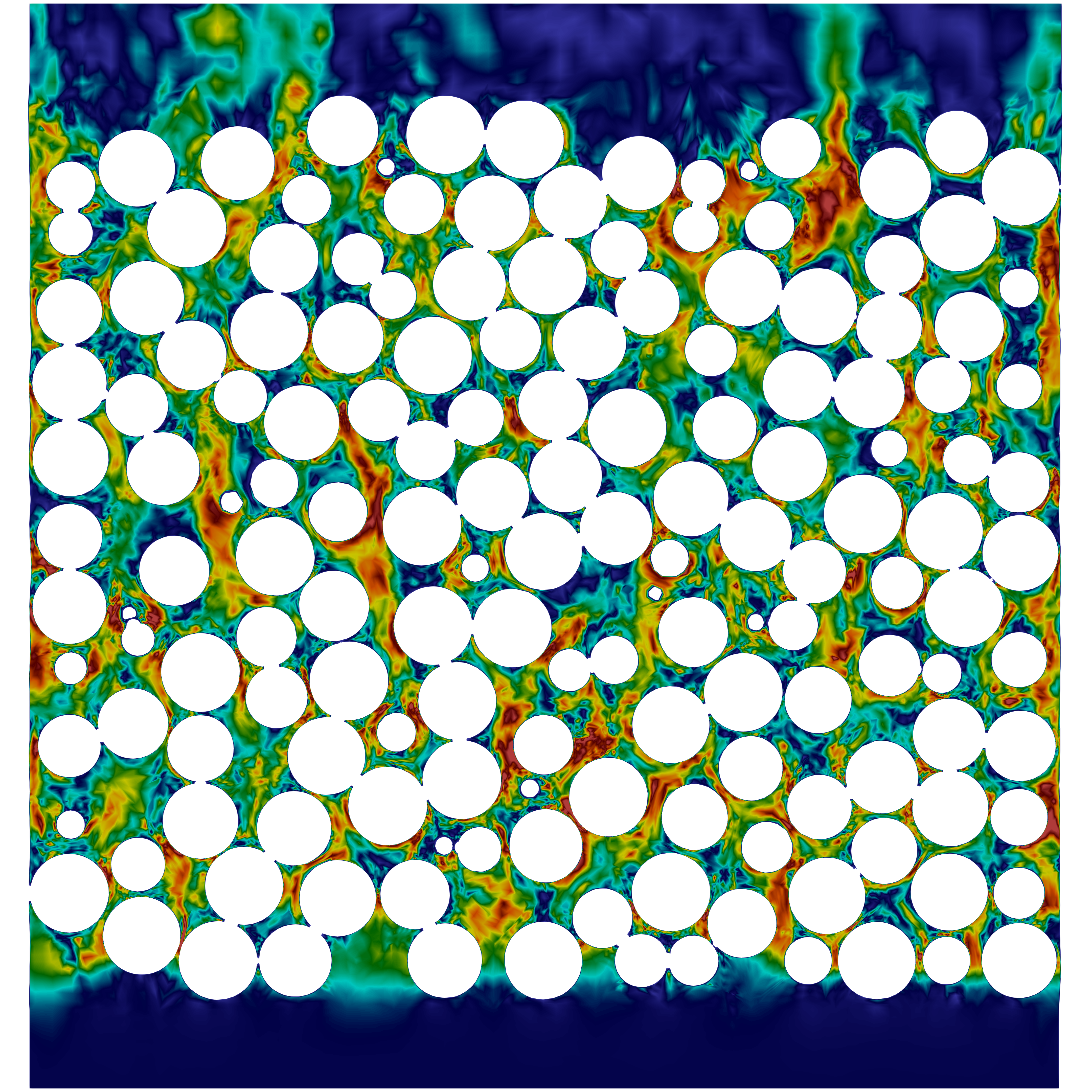}
        \caption{$P=2$ input}
        \label{fig:pb1568_p2}
    \end{subfigure}
    \hfill
    \begin{subfigure}[t]{0.53\linewidth}
        \centering
        \includegraphics[width=\linewidth]{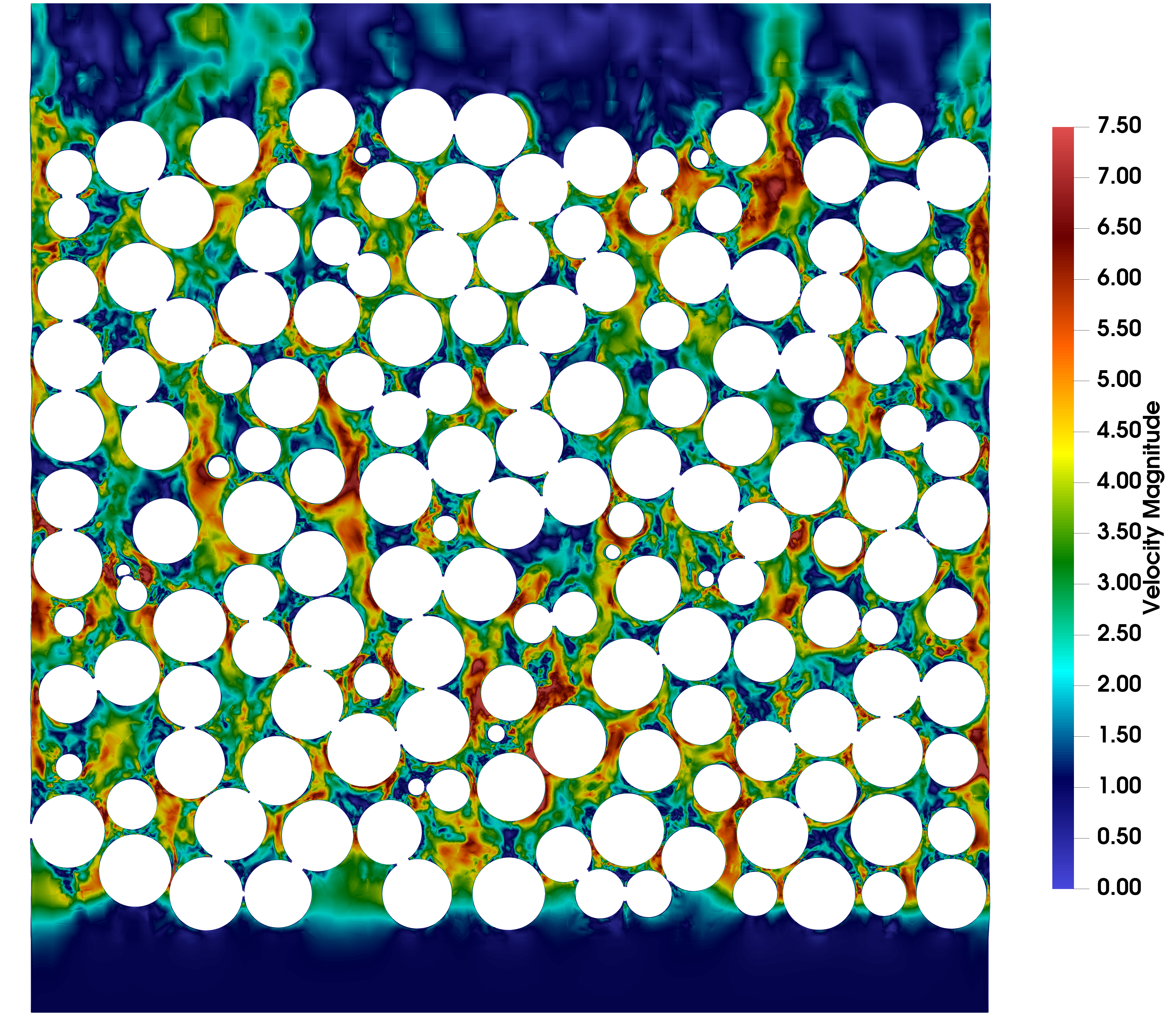}
        \caption{SR-GNN prediction}
        \label{fig:pb1568_pred}
    \end{subfigure}

    \caption{SR-GNN inference of the nondimensional velocity magnitude for the 1568-pebble bed at $Re=5000$. The lower-order $P=2$ field is used as input to the model, and the predicted field represents the reconstructed high-order velocity field.}
    \label{fig:pb1568_inference}
\end{figure}

The SR-GNN workflow was also applied to a larger packed-bed geometry containing 1568 pebbles. This case is used as a preliminary assessment of the applicability of the trained model to a significantly larger geometry than the 146-pebble bed used for the main restart assessment. Since the clearest restart improvement in the 146-pebble case was observed at $Re=5000$, the present large-geometry demonstration focuses on this Reynolds number.

\begin{figure}[h!]
    \centering

    \begin{subfigure}[t]{0.49\linewidth}
        \centering
        \includegraphics[width=\linewidth]{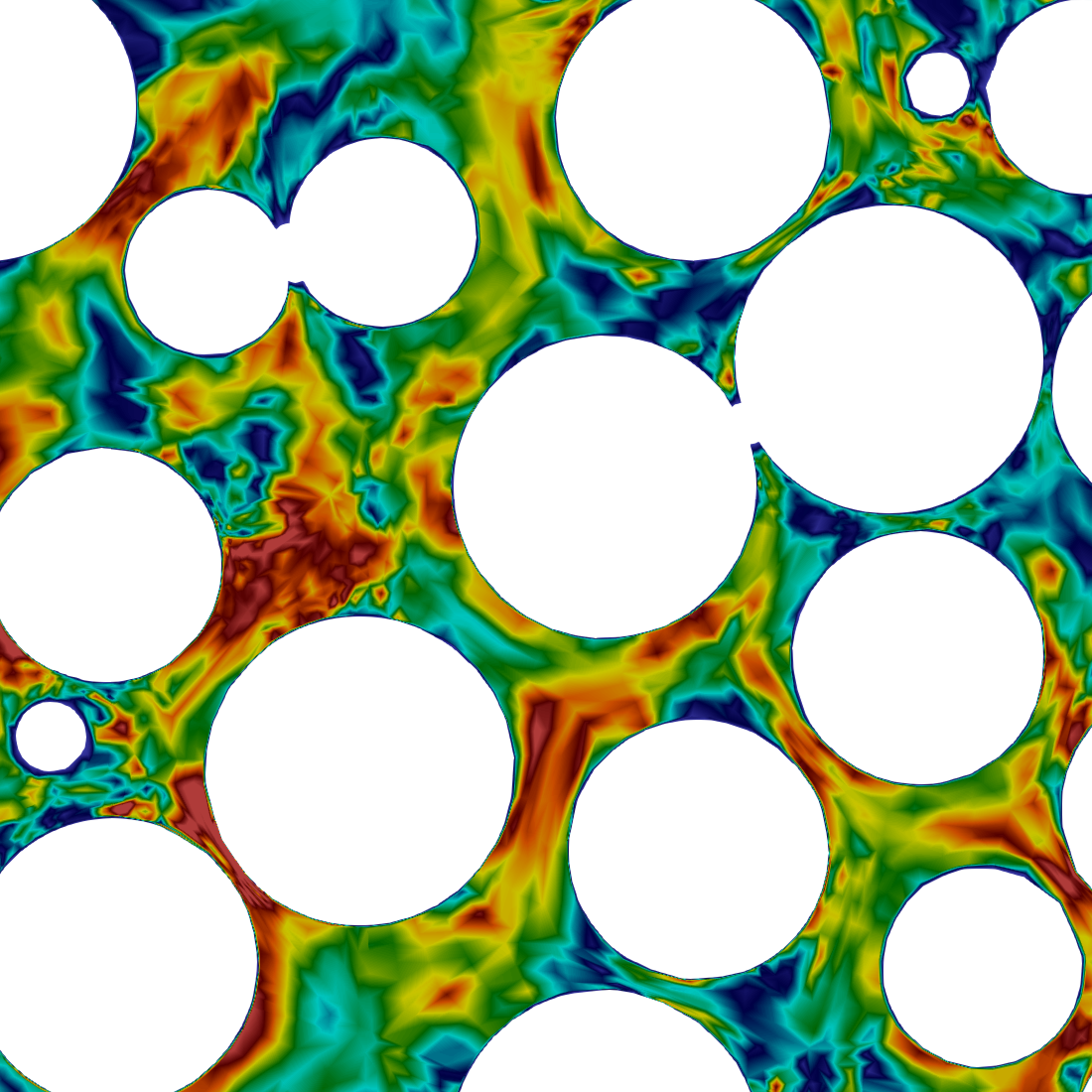}
        \caption{$P=2$ input}
        \label{fig:pb1568_p2_zoom}
    \end{subfigure}
    \hfill
    \begin{subfigure}[t]{0.49\linewidth}
        \centering
        \includegraphics[width=\linewidth]{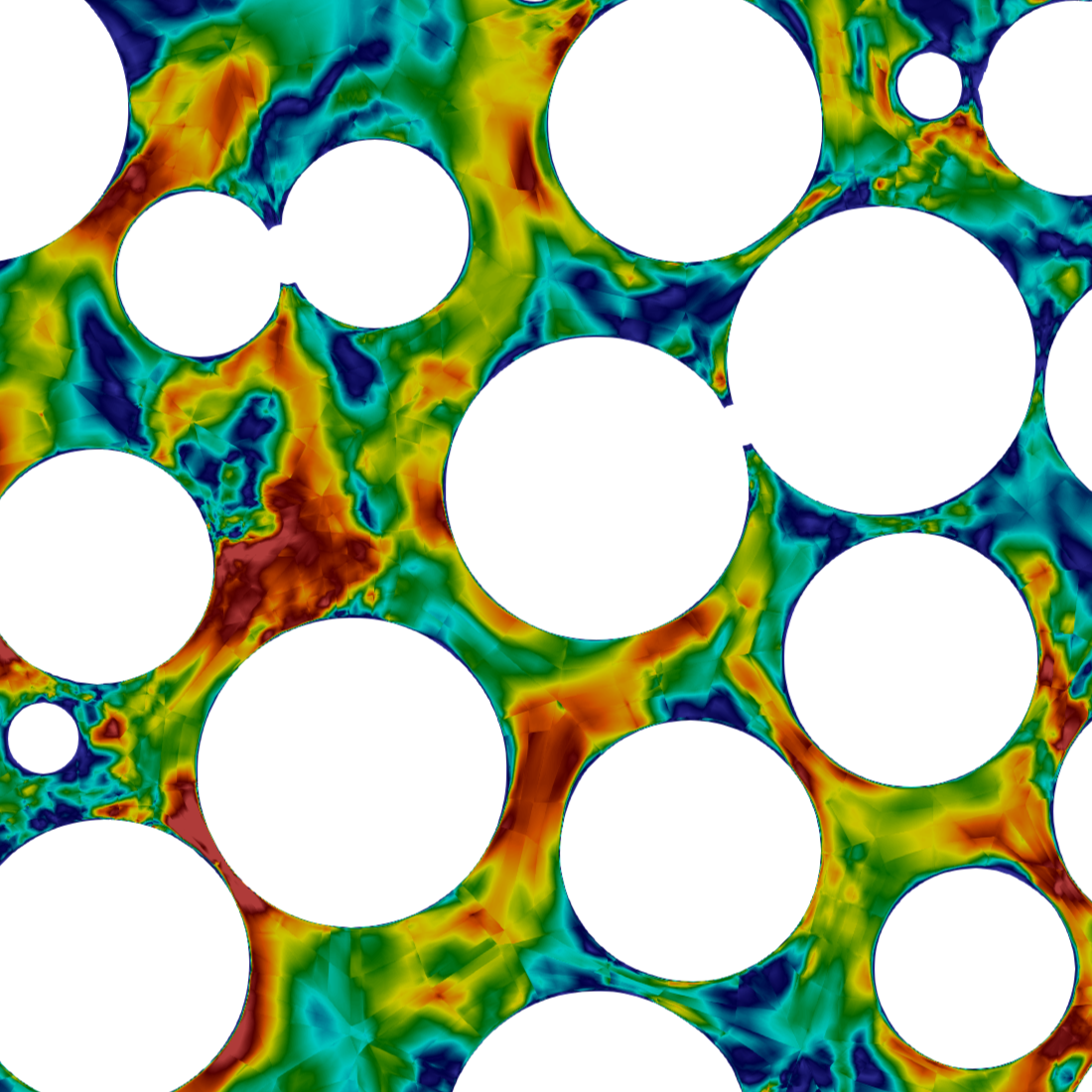}
        \caption{SR-GNN prediction}
        \label{fig:pb1568_pred_zoom}
    \end{subfigure}

    \caption{Zoomed view of the nondimensional velocity magnitude for the 1568-pebble bed at $Re=5000$. The reconstructed field shows additional fine-scale spatial variation in the interstitial region compared with the lower-order input.}
    \label{fig:pb1568_zoom}
\end{figure}

For this case, a true $P=2$ NekRS simulation was used as the lower-order input to the SR-GNN model trained at $Re=5000$. The corresponding $P=7$ simulation of the 1568-pebble bed would contain approximately 1 billion local GLL solution points, making this case substantially more expensive than the 146-pebble configuration. Therefore, the present analysis is limited to qualitative inference visualization, while high-order restart assessment of the pressure-drop evolution is left for future work.

Figure~\ref{fig:pb1568_inference} compares the lower-order $P=2$ input field and the SR-GNN reconstructed field for the 1568-pebble bed. The prediction introduces additional small-scale structures relative to the lower-order input while preserving the main flow paths through the packed bed. A closer view of a selected region is shown in Fig.~\ref{fig:pb1568_zoom}, where the reconstructed field displays finer spatial variation within the interstitial regions. These results indicate that the trained SR-GNN workflow can be applied to a much larger packed-bed geometry, although quantitative validation through high-order restart simulations is still required.

\section{Conclusions}

This work investigated the use of a Super-Resolution Graph Neural Network to improve the initialization of high-order NekRS simulations for turbulent flow in pebble-bed reactor geometries. The proposed workflow uses lower-order NekRS fields as inputs to the SR-GNN model, reconstructs a higher-order velocity representation, and uses the reconstructed field as an initial condition for subsequent high-order restart simulations. The objective is not to replace the high-fidelity CFD calculation, but to reduce the high-order flow-development time required to approach statistically stationary behavior.

The primary assessment was performed using a 146-pebble bed, where separate SR-GNN models were trained for $Re=1000$, $Re=2500$, and $Re=5000$. Qualitative inference comparisons showed that the SR-GNN models introduce additional small-scale spatial variation relative to the lower-order $P=2$ inputs while preserving the main interstitial flow structures. High-order restart simulations were then used to determine whether these reconstructed fields provide improved initial conditions in terms of pressure-drop convergence.

For $Re=1000$ and $Re=2500$, the pressure-drop histories obtained from the direct $P=2$ restart and the SR-GNN-initialized restart were nearly identical. This indicates that, for these Reynolds numbers, the super-resolved field did not provide a measurable improvement in the global hydraulic convergence of the high-order restart. In contrast, for $Re=5000$, the SR-GNN-initialized case approached the statistically stationary $P=7$ pressure-drop range faster than both the direct restart from the true $P=2$ field and the reference $P=7$ simulation initialized from a uniform velocity field. This result demonstrates that, under the highest Reynolds number considered, the SR-GNN reconstruction can provide a more effective initial condition for high-order pebble-bed simulations.

The trained SR-GNN workflow was also applied to a larger 1568-pebble geometry at $Re=5000$. In this case, the model was used to reconstruct a higher-order representation from a true $P=2$ simulation, demonstrating the applicability of the workflow to a significantly larger packed bed. Since a corresponding high-order restart assessment was not performed for this geometry, the 1568-pebble result should be interpreted as a preliminary large-geometry demonstration rather than a proof of acceleration.

Overall, the results show that SR-GNN-based initialization is a promising approach for reducing the flow-development cost of high-order CFD simulations in pebble-bed reactor geometries, particularly when the reconstructed field improves the global pressure-drop convergence. At the same time, the Reynolds number dependence observed in the restart results indicates that additional work is needed to improve robustness across flow regimes. Future work will focus on performing high-order restart assessments for the 1568-pebble bed, evaluating local velocity and turbulence statistics in addition to pressure drop, and training SR-GNN models using datasets that combine multiple Reynolds numbers and packed-bed geometries.

\section*{Acknowledgments}

This material is based upon work supported by the U.S. Department of Energy, Office of Science, under contract DE-AC02-06CH11357. An award of computer time was provided by the U.S. Department of Energy’s (DOE) Innovative and Novel Computational Impact on Theory and Experiment (INCITE) Program. This research used supporting resources at the Argonne and the Oak Ridge Leadership Computing Facilities. The Argonne Leadership Computing Facility at Argonne National Laboratory is supported by the Office of Science of the U.S. DOE under Contract No. DE-AC02-06CH11357. The Oak Ridge Leadership Computing Facility at the Oak Ridge National Laboratory is supported by the Office of Science of the U.S. DOE under Contract No. DE-AC05-00OR22725.


\bibliographystyle{ans.bst}
\bibliography{cited}

\end{document}